\documentclass[pra,twocolumn,showkeys]{revtex4-1}
\usepackage{amsfonts}
\usepackage{graphicx}
\usepackage{dcolumn}
\usepackage{bm}

\include{biblio}

\begin{document}

\title{Infrared absorption property of silicon carbide-silica nanocables synthesized by ethanol pyrolysis}
\author{Ryongjin Kim$^{1}$}
\author{Song-Jin Im$^{2,*}$}
\author{Ju-Myong Han$^{1}$}
\author{Yong-Hua Han$^{2}$}
\author{Tae-Hua Pak$^{1}$}
\author{Yong-Guk Choe$^{1}$}
\author{Sam-Hyok Choe$^{1}$}
\affiliation{$^{1}$Chair of Semiconductor, Department of Material Sciense, \textbf{Kim Il Sung} University, Pyongyang, DPR Korea}
\affiliation{$^{2}$Chair of Optics, Department of Physics, \textbf{Kim Il Sung} University, Pyongyang, DPR Korea
\\$^{*}$ sj.im@hotmail.com, ryongnam7@yahoo.com}

\begin{abstract}
A controllable synthesis method for SiC-$\mathrm{SiO}_{2}$ nanocables has been proposed. The diameter of SiC core and thickness of $\mathrm{SiO}_{2}$ shell were changed by adjusting the flow ratio between Ar dilution gas and ethanol precursor. With increasing the flow, the enhancement of 1137 $\mathrm{cm}^{-1}$ peak was observed from fourier transform infrared spectroscopy (FTIR) spectra. This peak is considered to be originated from a highly disordered surface structure of $\mathrm{SiO}_{2}$ shell  which was enhanced with increasing the flow. The FTIR spectra show the 910$\mathrm{cm}^{-1}$ peak which is attributed to surface phonon resonance in the nanostructure of SiC exited by p-polarized field component.
\end{abstract}
\keywords{Nanostructures, Fourier transform infrared spectroscopy (FTIR), Surface Phonon Resonance}

\maketitle

\section{Introduction}

The discovery of carbon nanotubes in 1991 \cite%
{Iijima_1991} and their unique properties differing from bulk material have stimulated the study of one-dimensional (1-D) nano-sized materials. Among them, silicon carbide nanocables have been focused as a potential candidate for light-emitting device, nano-reinforced composite materials, nanoelectronic devices and catalyst supports due to their outstanding properties such as wide band-gap, high hardness, excellent thermal conductivity, high electron mobility, high saturation drift velocity, good chemical inertness, etc \cite%
{WPQin_2003}. In general, the as-synthesized SiC nanowires contain amorphous $\mathrm{SiO}_{2}$ on their surfaces and have SiC-$\mathrm{SiO}_{2}$ core-shell structure. Recently, this structure has been attracting much attention owing to their potential applications for optoelectronic device and field emitter \cite%
{WMZhou_2006,RWu_2015}. For application of nano-materials, it is important to understand their several properties as well as to synthesize them in various forms.

Fourier transform infrared spectroscopy (FTIR) provides the important information about the several chemical bond states of the sample. The phonon states in nano-sized particle or wire differing from those in bulk will be reflected in infrared (IR) absorption spectra. But, to our knowledge, a few reports \cite%
{WWang_2007,HWShim_2007,KSenthil_2008,GFZou_2006,XLi_2011} related to IR absorption properties of SiC-$\mathrm{SiO}_{2}$ nanocables have been presented and most of them were restricted to confirm the characteristic band positions of Si-C and Si-O bonds. In our earlier study \cite%
{Ryongjin_2010}, we found a shoulder peak of Si−O antisymmetric stretching mode centered at 1130$\mathrm{cm}^{-1}$, and attributed it to the interface effect of the open structure of chainlike $\mathrm{SiO}_{2}$/SiC nanocables on the base of the comparison to the prior report \cite%
{QLHu_2003}. But, our further study showed that the shoulder peak of Si−O antisymmetric stretching mode could not be attributed to above mentioned origins. Furthermore, we found a novel peak centered at 910$\mathrm{cm}^{-1}$ from SiC-$\mathrm{SiO}_{2}$ nanocables in the IR absorption measurement. In this paper, we discuss the likely origin of these two peaks based on some experimental data and theoretical calculations.

\section{Experimental details}

SiC-$\mathrm{SiO}_{2}$ nanocables were synthesized by use of ethanol pyrolysis. Ethanol is thermally decomposed into CO, $\mathrm{H}_{2}$O, $\mathrm{C}_{2}$$\mathrm{H}_{6}$, C$\mathrm{H}_{4}$, $\mathrm{C}_{2}$$\mathrm{H}_{4}$, and $\mathrm{H}_{2}$  at about $700^{\circ}\mathrm{C}$ and the higher temperature above $1000^{\circ}\mathrm{C}$ induces the decomposition of hydrocarbon into C and $\mathrm{H}_{2}$ \cite%
{Ruiz_2007}. At about $1100^{\circ}\mathrm{C}$, the main products are CO and $\mathrm{H}_{2}$. Therefore, the use of thermal decomposition of ethanol allows us to obtain carbon source for synthesis of SiC-$\mathrm{SiO}_{2}$ nanocables at low cost \cite%
{Ryongjin_2010}. In our experiment, silicon source for synthesis SiC-$\mathrm{SiO}_{2}$ nanocables was generated by use of the reaction between Si(111) substrate (0.09$\Omega\cdot $cm, n-type) and pyrolysis productions of ethanol \cite%
{Ryongjin_2010, Ryongjin_2009}. 

The experimental details for synthesis of SiC-$\mathrm{SiO}_{2}$ nanocables were presented in our earlier report \cite%
{Ryongjin_2010}. We used five flow ratios between Ar and (Ar+ethanol), 98:2, 96:4, 94:6, 92:8, and 90:10, for synthesis of nanocables having different core diameters and shell thicknesses. After synthesis, the whole surface of the Si substrate was covered with white-colored product. Beyond the range of the flow ratios used in this experiment, the surface of Si substrate showed no more white color and changed from wheat to black in color \cite%
{Ryongjin_2010}. The synthesized products were characterized by scanning electron microscopy (SEM), high-resolution transmission electron microscopy (HRTEM), energy-dispersive X-ray diffraction (EDS), FTIR, and Raman spectroscopy. 

\section{Results and discussion}
 
\begin{figure*}
\includegraphics[width=1\textwidth]{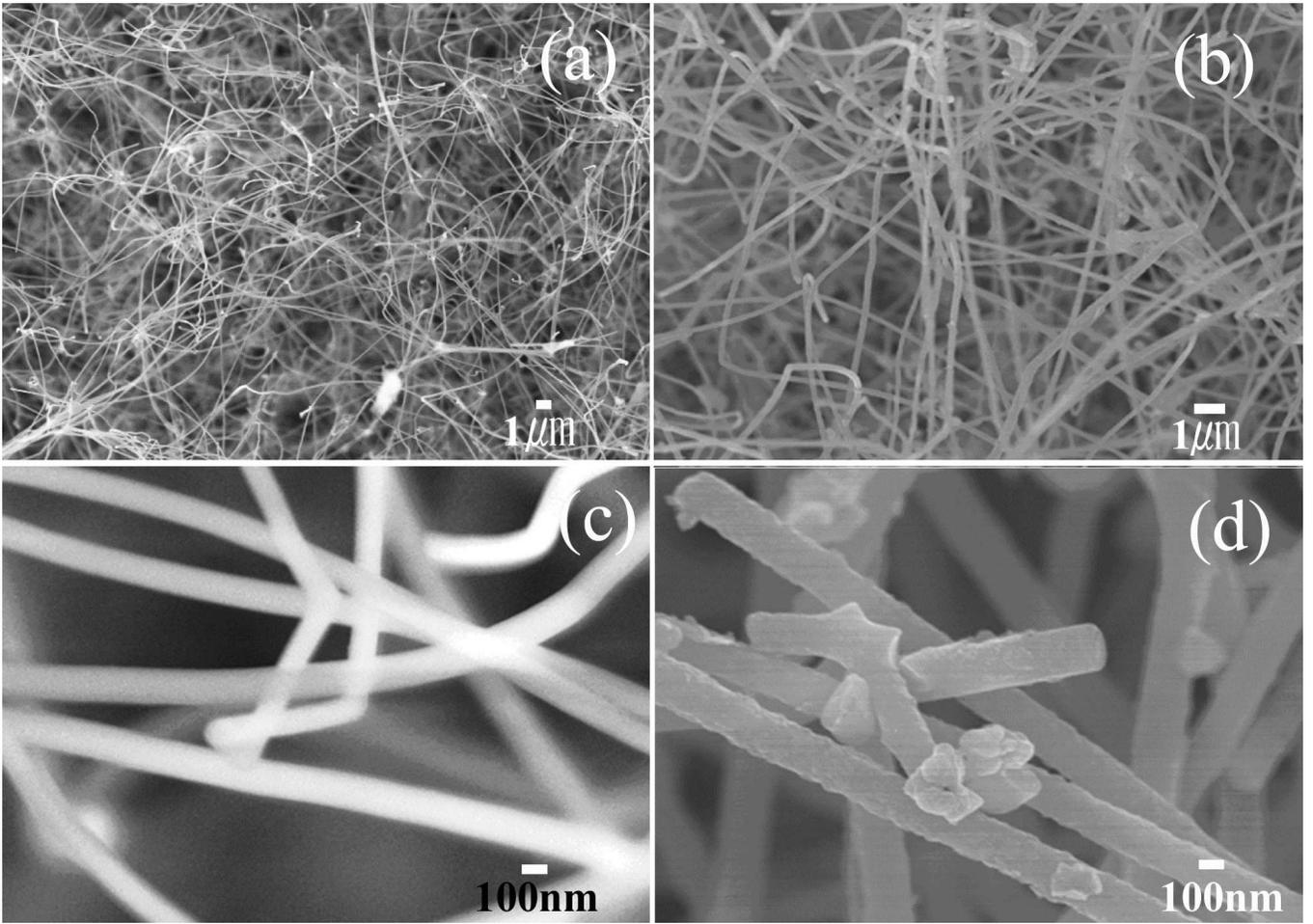}
\caption{SEM images of SiC nanocables synthesized at the flow ratios of 98:2 (a), (c) and 90:10 (b), (d).}
\label{fig:1}
\end{figure*}

SEM images of nanocables synthesized at the flow ratios of 98:2 Fig. \ref{fig:1}(a) and 90:10 Fig. \ref{fig:1}(a) were presented in Fig. \ref{fig:1}. Randomly oriented large-scale nanocables have been uniformly synthesized on the surface of Si substrate.
High-magnification SEM images show that the nanowires synthesized at 98:2 flow ratio have smooth surface and those synthesized at flow ratio of 90:10 have rough one. The mean diameter of nanowires synthesized at flow ratio of 98:2 and 90:10 were estimated to be about 140nm and 200nm, respectively. 
 
\begin{figure*}
\includegraphics[width=0.8\textwidth]{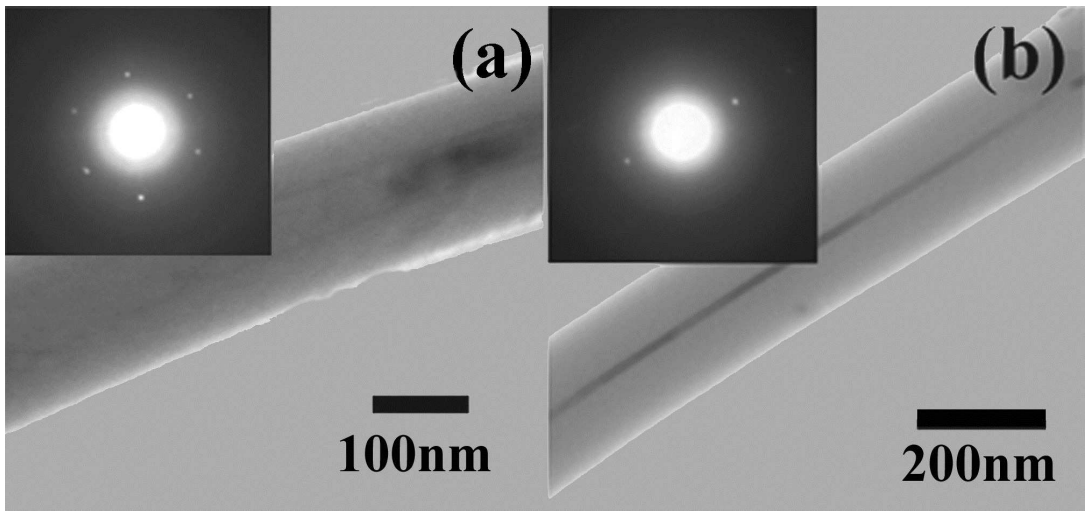}
\caption{TEM images of nanocables synthesized at flow ratios of 90:10 (a) and 98:2 (b).}
\label{fig:2}
\end{figure*}

Based on low-magnification TEM observation Fig. \ref{fig:2}(a) and \ref{fig:2}(b), the as-synthesized one dimensional nanostuructures have core-shell structure. With decreasing the ethanol flow, the thickness of shell increased and the core diameters decreased. For instant, mean diameter of cores and mean thickness of shells were estimated to be 40nm and 50 nm at 90:10 flow ratio, respectively. For those synthesized at 98:2 flow ratio, the values were estimated to be 10nm and 90 nm, respectively. Based on our previous reports \cite%
{Ryongjin_2010}, the nanocables synthesized by use of ethanol pyrolysis consisted of crystalline SiC cores and amorphous $\mathrm{SiO}_{2}$ shells. SAED (selected area electron diffraction) patterns [inserts in Fig. \ref{fig:2}(a) and (b)] recorded from the nanocables also confirm the existence of crystalline SiC and amorphous materials in nanocables. 

\begin{figure}
\includegraphics[width=0.4\textwidth]{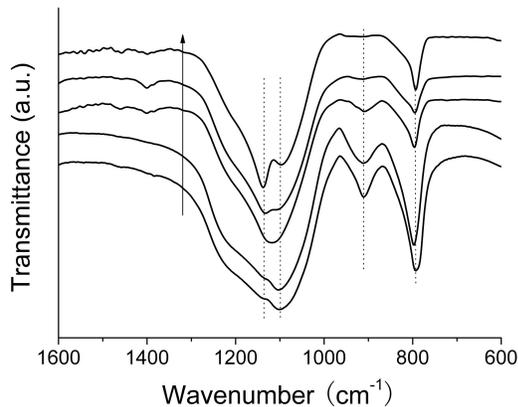}
\caption{FTIR spectra of nanocables synthesized at different flow ratios. Arrow indicates the increasing of ethanol flow.}
\label{fig:3}
\end{figure}

Fig. \ref{fig:3} shows the IR spectra of the nanocables synthesized at different flow ratios. The two strong absorption bands centered at $\sim$1090 and $\sim$796$\mathrm{cm}^{-1}$ in all spectra can be attributed to Si-O antisymmetric stretching mode \cite%
{LWLin_2007} and SiC transverse optical (TO) mode \cite%
{WWang_2007, HWShim_2007, KSenthil_2008, GFZou_2006}, respectively. Therefore the amorphous material is confirmed to be $\mathrm{SiO}_{2}$. 

It is interesting to note the change of Si−O main stretching band with ethanol flow. At low ethanol flow, Si−O main stretching band shows a strong absorption peak centered at 1090$\mathrm{cm}^{-1}$ and a weak shoulder centered at 1137$\mathrm{cm}^{-1}$. With increasing ethanol flow, the intensity of 1090$\mathrm{cm}^{-1}$ peak becomes weak, while 1137$\mathrm{cm}^{-1}$ shoulder peak is enhanced. At high ethanol flow, the two bands are clearly distinguished and the 1137$\mathrm{cm}^{-1}$ peak becomes stronger.

\begin{figure*}
\includegraphics[width=0.8\textwidth]{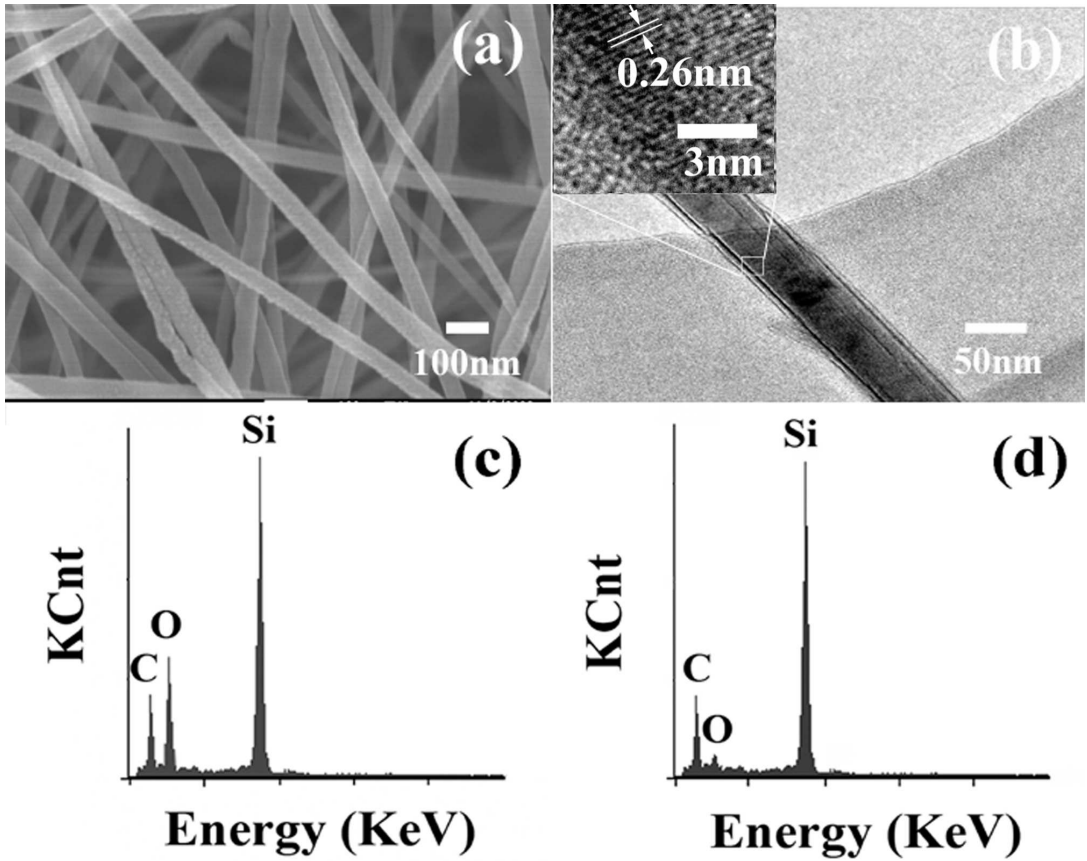}
\caption{SEM (a) and TEM (b) images of etched nanocables and EDS spectra before (c) and after (d) etching.}
\label{fig:4}
\end{figure*}

Similar shoulder peaks centered at about 1130$\mathrm{cm}^{-1}$ have been found from Si-$\mathrm{SiO}_{2}$ nanocables  \cite{QLHu_2003}, amorphous $\mathrm{SiO}_{2}$ nanowires \cite%
{LWLin_2007} and SiC-$\mathrm{SiO}_{2}$ nanocables \cite%
{Alan_2007}.

Hu et al.  \cite{QLHu_2003} considered that this shoulder peak from Si-$\mathrm{SiO}_{2}$ nanocables resulted from the interface effect of the open structure of chainlike $\mathrm{SiO}_{2}$/Si nanowires and the vibration of an interstitial oxygen atom in a silicon single-crystalline core of nanowire. For $\mathrm{SiO}_{2}$ nanowires \cite{LWLin_2007}, the origin of this shoulder peak was attributed to the structural disorder in amorphous $\mathrm{SiO}_{2}$. There was no explanation about this peak for SiC-$\mathrm{SiO}_{2}$ nanocables \cite{Alan_2007}. In previous reports \cite{QLHu_2003,LWLin_2007} the origin of this shoulder peak was interpreted on the base of the experimental result of Gaskell et al. \cite{PHGaskell_1976}. We had also found the shoulder peak from SiC-$\mathrm{SiO}_{2}$ nanocables \cite{Ryongjin_2009} and attributed it to the interface effect of the open structure of chainlike $\mathrm{SiO}_{2}$/SiC nanocables and the vibration of interstitial oxygen atoms in a silicon carbide single-crystalline cores of nanocables on the base of prior reports \cite{QLHu_2003}.
 
\begin{figure}
 \includegraphics[width=0.4\textwidth]{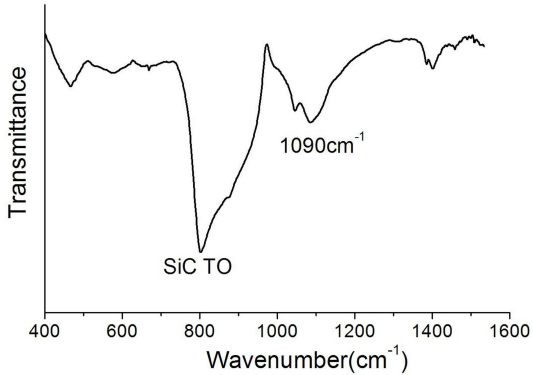}
 \caption{FTIR spectrum of etched nanocables.}
 \label{fig:5}
 \end{figure}

However, our further study on the IR absorption property of SiC-$\mathrm{SiO}_{2}$ nanocables made us doubt with the attributions of 1130$\mathrm{cm}^{-1}$ shoulder peak. For instant, the open structure didn't exist in $\mathrm{SiO}_{2}$ nanowires showing intensive and sharp 1130$\mathrm{cm}^{-1}$ shoulder peak \cite{LWLin_2007}. Although the sharp 1130$\mathrm{cm}^{-1}$ peak observed from $\mathrm{SiO}_{2}$ nanowires having mean diameter of 100nm  \cite{QLHu_2003} but didn't appear from $\mathrm{SiO}_{2}$ nanoparticles having diameters ranged 40-60nm  \cite{QLHu_2003}. Furthermore, 1130$\mathrm{cm}^{-1}$ shoulder peak was disappeared after 5\%HF etching for 2 min in spite of the existence of SiC/$\mathrm{SiO}_{2}$ interface as shown in Fig. \ref{fig:5} With respect to the origin of 1130$\mathrm{cm}^{-1}$ peak, we also believe the experimental result performed by Gaskell et al. \cite{PHGaskell_1976} and refer it to the structural disorder in amorphous $\mathrm{SiO}_{2}$ shells. The problem is why the structural disorder is created and where the structural disorder in $\mathrm{SiO}_{2}$ exists. The SEM, TEM and EDS measurements (Fig. \ref{fig:4}) before and after etching obviously identify the removal of amorphous $\mathrm{SiO}_{2}$ shell from nanocables during etching. These results and comparative analysis of prior reports imply that the structural disorder occurs near the surface of $\mathrm{SiO}_{2}$ and its enhancement depends on the synthesis condition. The rough surface [Fig. \ref{fig:1}(d)] implies the structural disorder of $\mathrm{SiO}_{2}$ surface. At this time, it is not clear why the shortening of Si-O bond occur in the surface of $\mathrm{SiO}_{2}$ shell of SiC-$\mathrm{SiO}_{2}$ nanocable. It needs further study. 

Next interest comes from a sharp band ranged from 870$\mathrm{cm}^{-1}$ to 960 $\mathrm{cm}^{-1}$ in Fig. 3. At low ethanol flow, an obvious peak centered at 910 $\mathrm{cm}^{-1}$ comes into view. With increasing ethanol flow, this peak becomes weak and finally a broad and weak absorption band appears in this region. To our knowledge, there is no report with respect to the sharp 910$\mathrm{cm}^{-1}$ peak in SiC-$\mathrm{SiO}_{2}$ nanocables. This peak is difficult to be attributed to the terminal Si−OH deformation band absorption having its maximum position at 958$\mathrm{cm}^{-1}$ \cite%
{LWLin_2007} due to the large frequency difference. Moreover, this peak decreased with increasing ethanol flow. If we assign this peak to the OH group, the intensity of this band should be increased with increasing ethanol flow because more OH groups would be formed on the surface of $\mathrm{SiO}_{2}$ shell at high ethanol concentration. This peak also differs from the SiC IF (interface) modes often observed in Raman measurement \cite%
{YYan_2003}, because no dominant peak appears in this range from etched nanocables as shown in Fig. 5.
 
\begin{figure}
\includegraphics[width=0.4\textwidth]{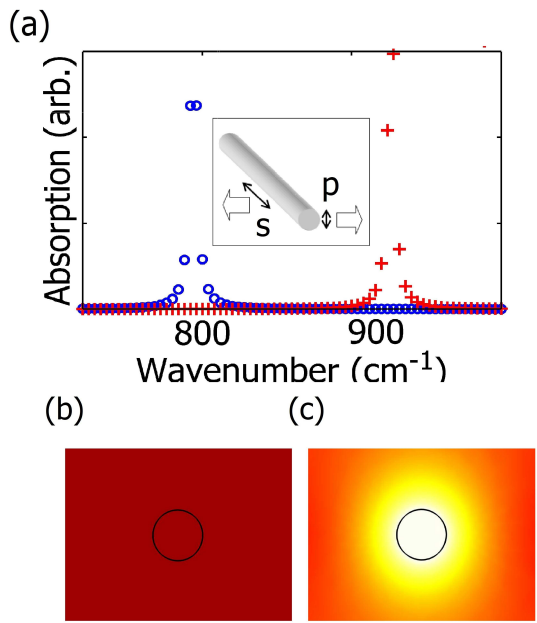}
\caption{Simulated results of infrared absorption in a SiC nanorod inserted into a transparent host medium. Infrared absorption spectra for incident fields with the s- (blue circle) and p- (red cross) polarizations (a). Field distributions for s- (b) and p- (c) polarized incident fields.}
\label{fig:6}
\end{figure}

To find out nature of the 910$\mathrm{cm}^{-1}$ peak, we simulate infrared field in a SiC nanorod inserted into a transparent host medium. Absorption in the SiC nanorod is calculated by $\vec{J}\cdot\vec{E}=\omega{}$Im$\left[\varepsilon(\omega)\right]\langle \vec{E}^{2}\rangle$ \cite%
{John_1999}, where $\langle\rangle$ signifies field-averaging over one optical cycle,  Im signifies imaginary part, $\omega$ is the frequency of interest, $\varepsilon(\omega)$ is the infrared permittivity of SiC at the frequency $\omega$, $\vec{J}$ is the displacement current density, and   $\vec{E}$ is the electric field strength. Fig. \ref{fig:6}(a) shows infrared absorption spectra for incident fields with the s- (blue circle) and p- (red cross) polarizations. For the s-polarization perpendicular to the cross section of SiC nanorod, there is one peak near 795$\mathrm{cm}^{-1}$ obviously corresponding to TO-mode of SiC, which is naturally observed in bulk SiC. For the p-polarization parallel to the cross section of SiC nanorod, there is one peak between TO-frequency and LO-frequency of SiC. This peak is attributed to resonant coupling of p-polarized field to the nanostructure of SiC, corresponding to surface phonon resonance (SPR) in the SiC nanorod. Fig. \ref{fig:6}(c) shows field enhancement in the SiC nanorod at the surface phonon resonance, while (b) shows no field enhancement at the TO-frequency for the s-polarization. 
An analytical treatment of an ellipsoid in the electrostatic approximation \cite%
{Craig_1983} can make nature of the two peaks more clear. The analytical study leads to the following expression for the polarizabilities  $\alpha_{i}$ along the principal axes ($i=1, 2, 3$)
\begin{eqnarray}
\alpha_{i}=4\pi{}a_{1}a_{2}a_{3}\frac{\varepsilon(\omega)-\varepsilon_{h}}{3\varepsilon_{h}+3L_{i}\left[\varepsilon(\omega)-\varepsilon_{h}\right]}.
\end{eqnarray} 
Here $a_{1},a_{2},a_{3}$ is semiaxes of the ellipsoid, $\varepsilon_{h}$ is the permittivity of host medium, $L_{i}$ is a geometrical factor given by
 \begin{eqnarray}
 L_{i}=\frac{a_{1}a_{2}a_{3}}{2}\int_{0}^{\infty}\frac{dq}{(a_{i}^{2}+q)}f(q),
 \end{eqnarray}
 \begin{eqnarray}
 f(q)=\sqrt{(q+a_{1}^{2})(q+a_{2}^{2})(q+a_{3}^{2})}
 \end{eqnarray}
 The geometrical factors satisfy $\sum{L_{i}}=1$, and for a sphere $L_{1}=L_{2}=L_{3}=\frac{1}{3}$. If the nanorod is considered as a kind of ellipsoid with $a_{1}=a_{2}$, $a_{3}=\infty$, $L_{1}=L_{2}=\frac{1}{2}, L_{3}=0$. If we remind the resonance arises at a pole of the polarizability, a surface-resonance peak ($L_{1}=L_{2}=L_{3}$) centered between the TO-frequency and the LO-frequency and no peak exactly centered at the TO-frequency should be observed in FTIR spectra of spherical nanoparticles of SiC as were in \cite%
{Sasaki_1989,GWMeng_2000}. However, for the case of nanorod, both of the surface-resonance peak ($L_{1}=L_{2}$) between the TO-frequency and the LO- frequency and the bulk-peak at the TO-frequency ($L_{3}=0$) should be observed as were clearly in our FTIR spectra. 

Our experimental result (Fig. \ref{fig:3}) shows weakening and disappearing of the 910$\mathrm{cm}^{-1}$ band with increasing the flow of ethanol precursor, contrary to the theoretical deduction. It is assumed that the discrepancy is attributed to the highly disordered surface of $\mathrm{SiO}_{2}$ shell and a transient layer between the $\mathrm{SiO}_{2}$ shell and the SiC core, which would be enhanced with increasing the flow of ethanol precursor, suppressing coupling of incident infrared field to the SiC core.

\section{Conclusion}

The enhancement of a peak centered at 1130$\mathrm{cm}^{-1}$ at higher frequency of Si-O stretching band was observed. The origin of this peak was referred to the structural disordering of $\mathrm{SiO}_{2}$ shell surface. Below 1000$\mathrm{cm}^{-1}$, our nanostructure reveals in its FTIR spectra both of the peak centered at 910$\mathrm{cm}^{-1}$ and the peak centered at 795$\mathrm{cm}^{-1}$. The 910$\mathrm{cm}^{-1}$ peak is attributed to surface phonon resonance in the SiC part which is characteristic of SiC nanoparticles. The 795$\mathrm{cm}^{-1}$ peak is attributed to TO-mode absorption in the SiC part, which is characteristic of bulk SiC.

\end{document}